\documentclass[manuscript,superscriptaddress]{revtex4}
\usepackage{amsfonts}
\usepackage{mathrsfs}
\usepackage{amsmath}
\usepackage{amssymb}
\usepackage{bm}

\usepackage{feynmp-auto}
\usepackage[normalem]{ulem}
\usepackage{extarrows}
\usepackage{slashed}
\usepackage{isodateo}
\usepackage{graphicx}
\usepackage{xcolor}
\usepackage[CJKbookmarks=true, bookmarksnumbered=true,bookmarksopen=true]{hyperref}
 \hypersetup{colorlinks,%
             linkcolor=[rgb]{0,0.2,0.6}, %
             citecolor=[rgb]{0,0.2,0.6}}

\usepackage{indentfirst}

\graphicspath{{fig/}}

\renewcommand{\FR}[2]{\displaystyle\frac{\,{#1}\,}{#2}}

\newcommand{\n}{\nonumber}

\def\bge{\begin{equation}}
\def\ede{\end{equation}}
\def\bga{\begin{aligned}}
\def\eda{\end{aligned}}
\def\bgp{\begin{pmatrix}}
\def\edp{\end{pmatrix}}
\def\bgs{\begin{subequations}}
\def\eds{\end{subequations}}
\newcommand{\order}[1]{\mathcal{O}({#1})}
\def\di{{\mathrm{d}}}
\def\D{{\mathrm{D}}}

\def\mb{\mathbf}

\def\pd{\partial}
\def\ld{{\mathscr{L}}}

\def\la{\langle}\def\ra{\rangle}
\def\sla{\slashed}

\def\to{\rightarrow}

\def\al{\alpha}

\def\de{\delta}

\def\lam{\lambda}

\def\Mp{M_{\text{Pl}}}

\newcommand{\ob}[1]{\mkern 2mu \overline{\mkern -2mu #1 \mkern -2mu}\mkern 2mu}

\begin{document}

\title{Standard Model Background of the Cosmological Collider}

\author{Xingang Chen}
\affiliation{Institute for Theory and Computation, Harvard-Smithsonian Center for Astrophysics, Cambridge, MA 02138, USA}

\author{Yi Wang}
\affiliation{Department of Physics, The Hong Kong University of Science and Technology, Hong Kong, P.R.China}

\author{Zhong-Zhi Xianyu}
\affiliation{Center of Mathematical Sciences and Applications, Harvard University, Cambridge, MA 02138, USA}

\begin{abstract}

The inflationary universe can be viewed as a ``Cosmological Collider'' with energy of Hubble scale, producing very massive particles and recording their characteristic signals in primordial non-Gaussianities. To utilize this collider to explore any new physics at very high scales, it is a prerequisite to understand the background signals from the particle physics Standard Model. In this paper we describe the Standard Model background of the Cosmological Collider.
\end{abstract}

\pacs{98.80.Cq, 04.62.+v}

\maketitle


Over more than 30 years since it was firstly introduced, the Cosmic Inflation \cite{Guth:1980zm} has been and still is the most promising paradigm describing the universe at its very early stage. It asserts that the universe has experienced a period of near exponential expansion with Hubble parameter $H$ up to $\order{10^{14}\text{GeV}}$, and the quantum fluctuations of spacetime during this period have seeded the density perturbations that in turn shape the large scale structure of the universe as we see today.

The ongoing observations of cosmological microwave background and large scale structure have achieved impressive precision, from which valuable information about primordial density perturbations can be extracted. For example, at the linear level, the scalar tilt of power spectrum has been measured with less-than-1\% accuracy \cite{Ade:2015xua}, together with the upper bound on the tensor-to-scalar ratio \cite{Array:2015xqh}, this has put nontrivial constraint on inflation models. Going beyond the linear level, a vast number of potential observables are available in the form of non-Gaussianities of primordial fluctuations, of which the simplest type is the bispectrum $S(k_1,k_2,k_3)$, coming from the 3-point correlation function $\la\zeta_{k_1}\zeta_{k_2}\zeta_{k_3}\ra$ of curvature perturbation $\zeta_k$ (or, similarly, of tensor perturbation).

The bispectrum contains information about interactions of inflaton fields among themselves and with other fields. Due to the high energy scale of inflation, particles with mass up to $\order{H}$ can be produced through quantum fluctuations. If such fields couple to inflaton, they can leave observable signal on the bispectrum of inflaton \cite{Chen:2009we,Arkani-Hamed:2015bza,Sefusatti:2012ye}. This process strongly resembles the workings of a particle collider. The cosmic expansion itself plays the role of an accelerator, while the final states of particle collision can be recorded in the non-Gaussianities, such as bispectrum $S(k_1,k_2,k_3)$, etc.

From the bispectrum $S(k_1,k_2,k_3)$ we may be able to extract useful information about massive particles $\Phi$ created during inflation, due to its interaction with the inflaton field $\phi$. This ``discovering channel'' of new physics is the so-called squeezed limit of $S(k_1,k_2,k_3)$, i.e. $k_1\simeq k_2\gg k_3$. In this channel, an intermediate particle of mass $m$ and spin $s$ will generate a characteristic signal of the following form \cite{Chen:2009we,Arkani-Hamed:2015bza},
\bge
\label{sqbispec}
  S(k_1,k_2,k_3)\!\propto C(\mu_s)\Big(\FR{k_3}{k_1}\Big)^{\mu_s}\!\! P_s(\cos\theta)+(\mu_s\!\!\to\!-\mu_s),
\ede
where $\mu_s$ is a constant dependent on both $m$ and $s$, and $\theta$ is the angle between $\vec k_2$ and $\vec k_3$. Taking $s=0$ for an example, we have $\mu_0=\sqrt{9/4-(m/H)^2}$. The coefficient $C(\mu_s)$ is of $\order{1}$ when $m\sim \order{H}$ but goes like $e^{-\pi m/H}$ when $m\gg H$. Consequently, a scalar with $m>\frac{3}{2}H$ ($m<\frac{3}{2}H$) will generate oscillatory (power-law) signals characterized by the imaginary (real) part of $\mu_s$.

One may attempt to conclude from the above discussion that the detection of a signal in form of eq.\;(1) in the squeezed limit of bispectrum would imply the existence of a new particle with mass $m\sim \order{H}$, and this may constitute an ideal explorer of new physics, given the fact that the inflation scale is typically much larger than electroweak scale, i.e., the Cosmological Collider typically has much higher energy than any real particle collider one would ever dream of.

However, it would be premature to attribute any signal like (\ref{sqbispec}) to new physics at scale $H$. In ordinary particle collider, it is extremely important to study the background signals from known physics, before we can claim any discovery of new physics. The same reasoning applies also to the Cosmological Collider, where the known physics, i.e. the Standard Model (SM) of particle physics, can generate interesting background. A careful study of this SM background would be the prerequisite for using the Cosmological Collider to explore any new physics. Any observational signal that deviates from this background would then be a sign of physics beyond the SM.

In this work we describe the SM background in the Cosmological Collider based on generic single field inflation, without specifying the inflation models. This task consists of two steps. The first step is to work out the SM spectrum during inflation, which turns out to be dramatically different from the usual one in the electroweak broken phase. The second one is to figure out how the SM fields enter the bispectrum. In both steps one needs a careful treatment of loop corrections in inflation background, which is both technically difficult and conceptually subtle. In this paper we focus on the physical interpretation of these results, and leave the detailed analysis of loop corrections to a companion publication \cite{CWX:toappear}.

\emph{SM Spectrum during Inflation. ---}
During inflation, the mass spectrum of SM depends not only on the background value of Higgs field, but also on the gravitational coupling and inflaton coupling of SM fields. The quantum corrections turn out to be crucial, too. To take account of all these factors more concretely, we assume the SM fields couple to gravity minimally so that all higher order gravitational couplings are suppressed and can be ignored, except for the unique dim-4 nonminimal coupling \cite{Bezrukov:2007ep,Espinosa:2015qea} between the Higgs field and Ricci scalar $\xi\sqrt{-g} R\mb H^\dag \mb H$ which can \textit{a priori} be large \cite{XRH2013}. We also assume generic single field slow-roll inflation models with the inflaton field $\phi$, and assume that the inflaton couples to SM fields through the following terms in the Lagrangian,
\begin{align}
\label{InfSMCoup}
\ld \supset\sum_\al\,f_\al(X,\phi)\mathcal{O}_\al[\Phi],
\end{align}
where $X\equiv(\pd_\mu\phi)^2$, $\mathcal{O}_\al[\Phi]$ represents any operator made from SM fields, denoted collectively by $\Phi$, and $f_\al(X,\phi)$ are  functions of $X,\phi$ which serve as a general parameterization of inflation-SM coupling. If the inflaton is SM gauge singlet, then so is $\mathcal{O}_\al$. But it is possible that the inflaton carries some SM charge, such as the case of Higgs inflation \cite{Bezrukov:2007ep}. Furthermore, we may also expect $f_\al(X,\phi)$ to depend on $X$ only due to the shift symmetry of inflaton field. But this is not compulsory as the shift symmetry is not exact. One can also consider higher order derivatives such as $\pd^2\phi$, but we shall not pursue this case here. In general, interactions in (\ref{InfSMCoup}) can be either direct couplings or generated from a messenger sector much heavier than $H$. On the other hand, the case of a messenger sector lighter than $H$ is not captured by (\ref{InfSMCoup}) and is left for future study.

Since the Higgs background plays a crucial role in determining the SM spectrum, we should distinguish three broad classes of possibilities:

\begin{enumerate}
\item \emph{Low scale inflation}: The Hubble scale $H$ during inflation is much smaller than the electroweak  broken scale. The Higgs field has the usual vacuum expectation value (VEV) $v\simeq 246$GeV, and the SM spectrum during inflation is the same as in flat space.

\item \emph{Non-Higgs inflation}: The Hubble scale $H$ is much larger than the electroweak broken scale and the inflaton is not the SM Higgs, so that Higgs VEV remains zero during inflation \footnote{For previous studies of Higgs background during inflation, see, e.g., A.~De Simone, H.~Perrier and A.~Riotto, JCAP {\bf 1301}, 037 (2013); Y.~Watanabe and J.~Yokoyama, Phys.\ Rev.\ D {\bf 87}, 103524 (2013).}.

\item \emph{Higgs inflation}: The SM Higgs boson itself is the inflaton. In typical Higgs inflation models, the normalized Higgs field acquires huge (time-dependent) VEV during inflation, usually around Planck scale $\Mp\simeq 2.4\times 10^{18}$GeV.
\end{enumerate}

The Case 1 is least interesting as it implies that the Cosmological Collider would have much lower energy than ground-based colliders. We shall not consider this possibility from now on. On the other hand, the SM spectrum turns out to be very different in Cases 2 and 3, both of which are very interesting as the Cosmological Collider achieves its full superiority. Below we describe these two cases in more detail. Since the inflation scale in these two cases is much higher than the electroweak scale, it is safe to neglect the negative quadratic term in SM Higgs potential in the following. 

In non-Higgs inflation with $H\gg v$, the Higgs VEV remains zero during inflation, based on which one may naively expect that all SM fields remain massless. But this is not true. Due to the Gibbons-Hawking temperature $T=H/2\pi$, the variance of the Higgs VEV is no longer zero. Mean field theory already tells us that this would contribute masses of order $\order{H}$ to various particles. To get more precise spectrum, we need to go beyond mean field theory. The detailed computation was performed in \cite{CWX:toappear}, and here we outline the results. First, within our framework, the Higgs mass is affected at classical level by the following three operators,\begin{align}
\label{Hlag}
\ld\supset &-\xi R \mb H^\dag\mb H \n\\
&- f_{H}(X,\phi)\mb H^\dag \mb H-f_{DH}(X,\phi)\mb|\D_\mu \mb H|^2,
\end{align}
and the Higgs field acquires a tree-level mass, \bge
M_{H0}^2=\FR{12\xi H^2+f_{H}(X_0,\phi_0)}{1+f_{DH}(X_0,\phi_0)},
\ede
where $X_0=-\dot\phi_0^2$ is the background value of $X$. Since the value of nonminimal coupling $\xi$ and background values of $f$ functions are unknown, and $\xi$ can be a priori large, we should treat the tree-level Higgs mass $M_{H0}^2$ as a free parameter. Secondly, an important point to note is that quantum corrections can generate nonzero contribution to Higgs mass even when $M_{H0}^2=0$. This is largely due to the infrared divergence (or late-time growth in the context of inflation) of loop correction \cite{Burgess:2009bs,Chen:2016nrs}. As shown in \cite{CWX:toappear}, the loop-corrected Higgs mass is given by,
\bge
\label{HiggsMeff}
M_{H}^2=\sqrt{\FR{3\lam}{8\pi^2}}\FR{4\big[1-\sqrt{\pi} ze^{z^2}\text{Erfc}(z)\big]}{-2z+\sqrt{\pi}(1+2z^2)e^{z^2}\text{Erfc}(z)}H^2,
\ede
where $\lam$ is Higgs self-coupling in SM, $z\equiv\sqrt{2\pi^2/3\lam}(M_{H0}/H)^2$ and $\text{Erfc}(z)\equiv 2\pi^{-1/2}\int_z^\infty\di t\,e^{-t^2}$. The above result reduces to the following one when $M_{H0}^2=0$,
\bge
  M_{H}^2=\sqrt{\FR{6\lam}{\pi^3}}H^2.
\ede

 Another striking phenomenon in inflation is that the gauge fields can also acquire nonzero mass even the Higgs VEV $\la \mb H\ra$ remains zero. This is again due to the infrared divergence of loop corrections. They can be calculated either using the real time Schwinger-Keldysh (SK) formalism with the dynamical renormalization group resummation of infrared divergence \cite{Chen:2016nrs}, or Wick-rotating the spacetime to Euclidean dS \cite{CWX:toappear}. Both methods yield the same results, i.e., the gluon and photon still remain massless during inflation, while $W/Z$ bosons acquire nonzero mass,
\begin{align}
\label{MW}
 &M_{W}^2=\FR{3g^2H^4}{8\pi^2M_H^2},
 &&M_{Z}^2=M_W^2/\cos^2\theta_W,
\end{align}
where $g$ is the weak gauge coupling and $\theta_W$ is the Weinberg angle. On the other hand, the fermions always remain massless during inflation so long as the Higgs VEV is zero.

It should be noted that the coupling constant $g$ in (\ref{MW}) also receives corrections from the operator $-\frac{1}{4}f_{W}(X,\phi)W_{\mu\nu}^2$ coming from (\ref{InfSMCoup}), and the corrected $g$ is related to its SM value $g_{\text{SM}}$ via $g^2=g_{\text{SM}}^2/(1+f_{W}(X_0,\phi_0))$. Similar corrections apply to other SM couplings including the Weinberg angle. Generally it makes the SM background rather arbitrary, but tractable cases do exist when the background values of various $f_\al$ functions are small enough, and in such cases one can make certain predictions to SM spectrum.

With SM spectrum in non-Higgs inflation clarified, now we turn to typical Higgs inflation, where the Higgs field itself is the inflaton and acquires a huge VEV. Specificially, let's take the original Higgs inflation model as an example \footnote{Original Higgs inflation model suffers the problem of Higgs instability at inflation scale, but it can be relieved by modifying the model slightly by introducing new physics \cite{He:2014ora}.}. There the Higgs inflaton $\phi\simeq 5\Mp$ at the beginning of observable inflation. As a result, all charged fermions and $W/Z$ bosons acquire masses proportional to their Higgs coupling and also to the Higgs VEV, but with $v=246$GeV replaced by the quantity $\frac{\Mp}{\sqrt\xi}\big(1-e^{-\sqrt{2/3}\phi/\Mp}\big)^{1/2}\sim\order{0.01\Mp}$, where $\xi$ can be as large as $10^4$. On the other hand, since the Higgs field itself is the inflaton, its mass (effectively zero) does not belong to the isocurvaton spectrum.

\begin{figure}[t]
\centering
\includegraphics[width=0.5\textwidth]{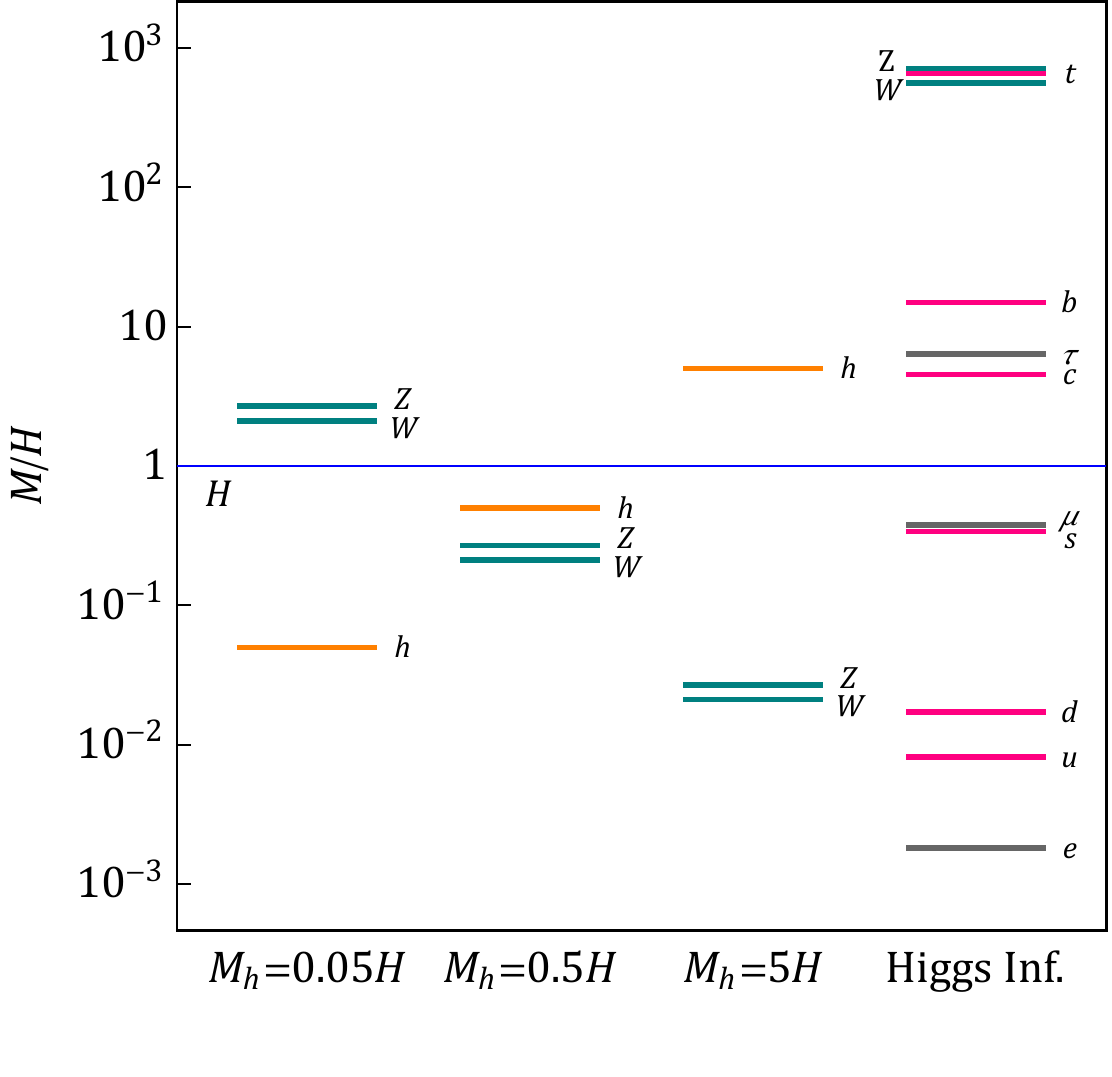}
\caption{The mass spectrum of Standard Model during inflation. The left three columns correspond the non-Higgs inflation with zero Higgs VEV, and Higgs masses are chosen to be $(0.05,0.5,5)H$, respectively. The rightmost column corresponds to the original Higgs inflation model.}
\label{fig_smsp}
\end{figure}

The main results of this section are summarized in Fig.\;\ref{fig_smsp}, where we plot the SM spectrum normalized by the Hubble scale $H$. For illustration, the Hubble scale is fixed to the value in typical Higgs inflation model, $H\simeq 2.0\times 10^{13}$GeV. For non-Higgs inflation cases (left 3 columns), all $f_\al$ functions are assumed to be negligibly small. All SM couplings are extrapolated to Hubble scale by 2-loop renormalization group running.

\begin{figure}[t]
\centering
\includegraphics[width=0.4\textwidth]{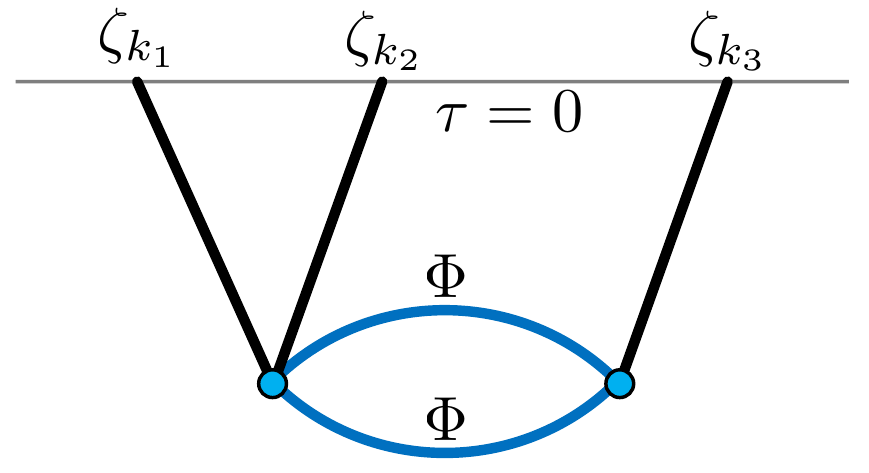}
\caption{The diagram contributing to the squeezed limit of the bispectrum with SM loop. The $\Phi$ field represents any of SM fields.}
\label{fig_smloop}
\end{figure}

\emph{SM Background in the Squeezed Limit. ---}
Given the SM spectrum during inflation, now we figure out the signals of SM fields in the inflaton bispectrum. All SM fields are charged under SM gauge group and thus are produced in pairs. Therefore, they contribute to 3-point inflaton correlators starting from 1-loop level. An example of this contribution is shown in Fig.\;\ref{fig_smloop}. There is also a 1-loop diagram with three 3-point vertices, but it is likely subdominant due to a further suppression factor $\dot\phi_0^2$.

It is important to specify the SM-inflaton couplings (\ref{InfSMCoup}) in order to evaluate diagrams such as in Fig.\;\ref{fig_smloop}. At 1-loop level, the only relevant operators $\mathcal{O}_\al$ are those quadratic in SM fields. SM couplings beyond quadratic order, such as Higgs self-coupling or gauge couplings, contribute to bispectrum at least through 2 loops, and thus will be neglected. If we further restrict our attention to scalar operators with dimension no greater than 4, then there are only 4 terms,
\begin{align}
\ld\supset& -f_{H}(X)\mb H^\dag \mb H-f_{DH}(X)\mb|\D_\mu \mb H|^2\n\\
&-f_{\Psi_i}(X)\ob{\Psi}_i\sla{\mathscr{D}}\Psi_i-\FR{1}{4} f_{A_a}(X)F_{a\mu\nu}F_a^{\mu\nu},
\end{align}
where $\Psi_i$ denotes all charged fermions and $\sla{\mathscr{D}}$ is corresponding covariant derivative, $F_{a\mu\nu}=\pd_\mu A_{a\nu}-\pd_\nu A_{a\mu}$ is the quadratic part of gauge kinetic term and $A_a$'s represent the SM gauge fields, and we have made a further simplifying assumption that various $f_\al$ functions depend only on $X$ but not directly on $\phi$, i.e. we are considering the leading terms under the shift symmetry. After separating the inflaton into background and fluctuation $\phi=\phi_0+\de\phi$, the operator $\mathcal{O}_\al\supset\{|\mb H|^2,|\D_\mu\mb H|^2,\ob{\Psi}_i\sla{\mathscr{D}}\Psi_i,F_{\mu\nu}^2\}$ couples to the inflaton fluctuation according to,
\begin{align}
  \ld\supset&~ f_{\al0}\mathcal{O}_\al+ 2f_{\al0}'\dot\phi_0\dot{\de\phi}\mathcal{O}_\al\n\\
  &~+\big[f_{\al0}'(\pd_\mu\de\phi)^2+2f''_{\al0}\dot\phi_0^2\dot{\de\phi}{}^2\big]\mathcal{O}_\al,
\end{align}
where $f'_{\al}=\di f_\al/\di X$, and the subscript 0 indicates that the background value has been taken. We can also drop the last term proportional to $\dot\phi_0^2$, which is expected to be much smaller than other terms.

Then we can apply the SK formalism to calculate the three-point correlation of $\de\phi$,
\begin{align}
\label{dephi3}
&\la\de\phi_{k_1}\de\phi_{k_2}\de\phi_{k_3}\ra'\n\\
=&~4f_{\al0}'^{2}\dot\phi_0
\int_{-\infty}^0\FR{\di\tau'\di\tau''}{(H^2\tau'\tau'')^2} \sum_{\text{SK}}\la\mathcal{O}_\al^2(k_3;\tau',\tau'')\ra'\n\\
&~\times\pd_{\tau''}G_{k_3}(0,\tau'')\big[ \pd_{\tau'}G_{k_1}(0,\tau')\pd_{\tau'}G_{k_2}(0,\tau')\n\\
&~+\mb k_1\cdot\mb k_2G_{k_1}(0,\tau')G_{k_2}(0,\tau')\big],
\end{align}
where the prime $\la\cdots\ra'$ indicates that the delta function of momentum conservation has been amputated, $G_k(\tau,\tau')$ is the SK propagator of inflaton with momentum $k$ from conformal time $\tau$ to $\tau'$, and the summation goes over all SK contours.

It is in general quite difficult to carry out the integral (\ref{dephi3}) in closed form. However, if we are only concerned with the ``non-local'' behavior as non-integer power of $k_3/k_1$ in the squeezed limit $k_{1,2}\gg k_3$, it is possible to get analytical expressions for (\ref{dephi3}), by expanding the correlator $\la\mathcal{O}_\al^2(k_3;\tau',\tau'')\ra'$ in the $\tau',\tau''\to 0$ limit. Remarkably, the result is free of UV divergence, and does present non-integer power of $k_3/k_1$. Here we present the results in terms of curvature perturbation $\zeta$, using the relation $\zeta=-H\delta\phi/\dot\phi_0$, and the standard parameterization,
\begin{align}
\label{zeta3}
\langle \zeta_{k_1}\zeta_{k_2}\zeta_{k_3}\rangle' \equiv
 \FR{(2\pi)^4}{(k_1k_2k_3)^2}P_{\zeta}^2S(k_1,k_2,k_3),
\end{align}
where $P_\zeta=H^2/(8\pi^2 \Mp^2 \epsilon)$ is the scalar power spectrum. Then, in the squeezed limit $k_S\equiv k_1\simeq k_2\gg k_3\equiv k_L$, the non-local part of (\ref{dephi3}) with $\mathcal{O}_\al\supset\{|\mb H|^2,|\D_\mu\mb H|^2,\ob{\Psi}_i\sla{\mathscr{D}}\Psi_i,F_{\mu\nu}^2\}$ can be collectively written in terms of $S(k_1,k_2,k_3)$ in (\ref{zeta3}) as,
\bge
\label{Salpha}
S_\al=\left\{
\begin{split}
&\mathcal{A}_\al\Big(\FR{k_L}{k_S}\Big)^{a_s-2\mu_s}+(\mu_s\to-\mu_s), \hspace{-7mm}&\mu_s\text{~real}\\
&2\text{Re}\bigg[\mathcal{A}_\al\Big(\FR{k_L}{k_S}\Big)^{a_s-2\mu_s}\bigg],&\mu_s\text{~complex}
\end{split}
\right.
\ede
where $\mu_s=\sqrt{b_s^2-(M_\al/H)^2}$, and $M_\al$ is the mass of the fields in $\mathcal{O}_\al$. The spin $s$-dependent parameter $a_s=(2,1,2)$ and $b_s=(\frac{3}{2},0,\frac{1}{2})$, for $s=(0,\frac{1}{2},1)$, respectively. The coefficient $A_\al$ depends on the choice of $\mathcal{O}_\al$. For dim-2 operator $|\mb H|^2$, this coefficient is given by $\mathcal{A}_{H}=f_{H0}'^2\dot\phi_0^2C_{H}(\mu_0)/\pi^4$, and $C_{H}(\mu_0)$ is a function of $\mu_0$ which is suppressed as $e^{-2\pi M_\al/H}$ when $M_\al\gg H$ but enhanced as $(M_\al/H)^{-4}$ when $M_\al\ll H$. On the other hand, for dim-4 operators $\mathcal{O}_\al\supset\{|\D_\mu\mb H|^2,\ob{\Psi}_i\sla{\mathscr{D}}\Psi_i,F_{\mu\nu}^2\}$, $\mathcal{A}_\al=f_{\al 0}'^2H^4\dot\phi_0^2C_\al(\mu_s)/\pi^4$, and the coefficient $C_\al(\mu_s)$ depends on $\mu_s$ only, and is again suppressed by $e^{-2\pi M_\al/H}$ when $M_\al\gg H$.

Some features of (\ref{Salpha}) are worth mentioning. Firstly, for sufficiently light particle, i.e. $m<b_s H$, the power $\mu_s$ is real and positive. In this case, the squeezed limit of bispectrum shows characteristic power-law behavior with (generally) non-integer exponent $\mu_s$. On the other hand, for heavy particle with mass $m>b_s H$, $\mu_s$ is imaginary and the corresponding bispectrum shows oscillatory behavior. Both power-law and oscillatory behaviors are distinctive signals of massive fields. Remarkably, massive fermions always show oscillatory signal rather than power-law.

Secondly, the Boltzmann suppression $e^{-2\pi m/H}$ appears in all cases, which means that we should not hope to observe particles with mass much larger than $H$.

Thirdly, we comment on the observability of this SM background. We have shown that the amplitudes of these specific bispectra are
$f_{NL} \sim f_{\al 0}'^2H^4\dot\phi_0^2C_\al(\mu_s)$ for dim-4 operators and a similar expression for dim-2 operators.
Given current experimental constraints, these bispetra are likely unobservable in CMB, but future experiments in large scale structure surveys \cite{Dore:2014cca} and the 21cm tomography \cite{Loeb:2003ya} are expected to significantly improve the constraints on primordial non-Gaussianities. For the type of signals we are interested in, it has been forecasted that future 21cm experiments can in principle be sensitive to $f_{NL}\gtrsim \order{0.01}$ \cite{21cm_forecast}.
Assuming $C_\al, C_H\sim\order{1}$, we see that the SM background would be detectable if $f_{\al 0}'^2\gtrsim (H^2\dot\phi_0)^{-1}$ and $f_{H0}'\gtrsim\dot\phi_0^{-1}$. This condition can be further loosen for sufficiently light bosons ($m<H/2$), because in this case the $C_\al$ factors for $|\mb H|^2$ and $F_{\mu\nu}^2$ operators are greatly enhanced and can be much larger than $\order{1}$.

Finally, we note an interesting window of parameter space where the SM background becomes both predictable and observable. This corresponds to region where $f_{\al 0}$ are sufficiently small but $f_{\al 0}'$ are sufficiently large. In this case the SM spectrum is not affected by the SM-inflaton couplings, and we may even hope to \emph{calibrate} the Cosmological Collider using this well-defined and distinct SM signal. This parameter range is likely unnatural in effective field theory, but it is interesting to investigate if this can be realized in concrete models. On the other hand, in the simplest case where $f_\alpha\sim X$, the amplitude of oscillatory/power-law signal in the bispectrum is unobservably small, so the SM background is negligible. For the Cosmological Collider program, both the observable and unobservable cases are important because they are the stepping stone to new physics beyond SM.

\emph{Summary and Discussions. ---}
In this Letter we have discussed the mass spectrum of SM during inflation and how this spectrum can be revealed in the squeezed limit of scalar bispectrum. The latter constitutes the background signal in the discovering channel of new physics at inflation scale on the Cosmological Collider. Loop corrections turn out to be crucial in determining both mass spectrum of SM and scalar bispectrum. 

An important lesson is that a detection of the power-law/oscillatory signal with apparent mass $m\sim H$ and spin $s$ does not necessarily imply a new heavy particle with measured mass and spin; it may also come from a known particle whose mass is affected by quantum correction or inflaton background. Similarly, starting from 1-loop order, the angular dependence in the bispectrum indicates only the total angular momentum of loop particles, rather than the intrinsic spin of a single particle. It should also be clarified that any detection of ``SM background'' is itself a sign of new physics, because the SM-inflaton coupling is most likely from a sector beyond SM. 

Many problems along this direction remain to be explored. A more systematic study of the SM background is needed, probably using a more general effective field theory formulation of inflation-SM system. At the same time, it is also desirable to work out the SM signals in concrete models of inflation, in particular in the ``calibration limit''. In the latter case one may also expect to discriminate inflation models using SM signals.

\begin{acknowledgments}
XC is supported in
part by the NSF grant PHY-1417421. YW is supported by grants HKUST4/CRF/13G and ECS 26300316 issued by the Research Grants Council (RGC) of Hong Kong. ZZX is supported in part by Center of Mathematical Sciences and Applications, Harvard University.
\end{acknowledgments}

\end{document}